# FACTORS INFLUENCING E-COMMERCE ADOPTION BY RETAILERS IN SAUDI ARABIA: A QUANTITATIVE ANALYSIS

Rayed AlGhamdi,
Griffith University, Australia
rayed.alghamdi@griffithuni.edu.au

Ann Nguyen
Griffith University, Australia
a.nguyen@griffith.edu.au

Jeremy Nguyen
Griffith University, Australia
j.nguyen@griffith.edu.au

Steve Drew
Griffith University, Australia
s.drew@griffith.edu.au

## ABSTRACT

This paper presents some findings from a study researching the diffusion and adoption of online retailing in Saudi Arabia. Although the country has the largest and fastest growing ICT marketplace in the Arab region, e-commerce activities have not progressed at a similar speed. In general, Saudi retailers have not responded actively to the global growth of online retailing. Accordingly new research has been conducted to identify and explore key issues that positively and negatively influence Saudi retailers in deciding whether to adopt the online channel. While the overall research project uses mixed methods, the focus of this paper is on a quantitative analysis of responses obtained from a survey of retailers in Saudi Arabia, with the design of the questionnaire instrument being based on the findings of a qualitative analysis reported in a previous paper. The main findings of the current analysis include a list of key factors that affect retailers' decision to adopt e-commerce, and quantitative indications of the relative strengths of the various relationships.

## 1. INTRODUCTION

The number of commercial organizations that tend to apply electronic commerce systems is on the rise. In the near future, this trend will become not only a tool to simply increase income, but without doubt will be considered an essential means for competition [1]. While developed nations have become familiar with e-commerce, e-commerce is still considered an innovation in the Saudi environment. Rogers [2] defined an innovation as an idea, practice or object that is perceived as new by an individual or other unit of adoption. Despite the fact that Saudi Arabia has the largest and fastest growth of ICT marketplaces in the Arab region [3], [4], [5], [6], e-commerce activities are not progressing at the same speed [7], [8], [9]. Tiny number of Saudi commercial organizations, mostly medium and large companies from the manufacturing sector and mainly Business to Business (B2B), are involved in e-commerce implementation [9]. The question which should be asked here, why retailers in Saudi Arabia are reserved adopting e-commerce?

In order to determine the most current factors influencing retailers in Saudi Arabia, a study was conducted that highlighted the perceptions of retailers. An exploratory research design was used and then followed in this paper using quantitative indications of the relative strengths of the various relationships. In the following section we review the global literature in terms of the factors affecting the adoption and use and present the contextual setting of government and business with regards to e-commerce as related in extant literature in Saudi Arabia. We then present a discussion of the Diffusion of Innovation model [2] and how it is applied here to the diffusion of e-retailing in Saudi Arabia. A description of an investigation methodology using a mix of qualitative and quantitative methods is presented followed by quantitative analysis of the results and a discussion of indicators to promote the growth of e-commerce within that country.

## 2. BACKGROUND AND PREVIOUS RESEARCH

Between 1995 and 2000, there was a notable proliferation to start-up/ adopt online retailing systems in USA [10], [11]. Since 2000, e-commerce activities rapid growth is obvious in the developed world. Global e-commerce spending has currently reached US$10 trillion and was US$0.27 trillion in 2000 [12]. The United States, followed by Europe, constitutes the largest share with about 79% of the global e-commerce revenue [12]. However, the African and Middle East regions have the smallest share with about 3% of the global e-commerce revenue [12].

Many studies have been conducted around the world to figure out challenges and drivers to online commerce. Nair [13] stated that among key factors that influence the development of e-commerce are ICT infrastructures including the rate of adoption of computers and maturity of Internet broadband, availability of online services such as online payment services, legislations and regulations, and security. A study was conducted a cross-country compression, covered USA, Brazil, Mexico, German, France, Denmark, China, Taiwan, Singapore and Japan, to "examine the key global, environmental and policy factors that act determinants of e-commerce diffusion" [14]. "It finds that B2B e-commerce seems to be driven by global forces, whereas B2C seems to be more local phenomena". The determinants which act as drivers/enablers for B2C e-commerce are (1) consumer desire for convenience, lifestyle enhancements, and greater product/service selection, especially among younger generation; (2) business desire to reach new markets or protect existing markets; (3) consumer purchasing power; (4) rapid Internet diffusion: high IT literacy, strong IT infrastructure; and (5) government promotion. In contrast, the determinants which act as barriers/inhibitors for B2C e-commerce are (1) lack of valuable and useful content for consumers; (2) inequality in socioeconomic levels; (3) consumer reluctance to buy online and lack of trust due to security/privacy concerns; (4) consumer reluctance to buy online due to preferences for in-store shopping; (5) existence of viable alternatives, such as dense retail networks, convenience stores; (6) lack of online payment mechanisms; (7) lack of customer service; (8) language differences.

"Government and industry promotion takes various forms from country to country, but the most common areas are promotion of IT and e-commerce in businesses, especially SMEs, by providing them with technical support, training, and funding for IT use" [14]. Government initiatives have had limited influences on the proliferation of e-commerce in most countries (which have been studied based on a 2002 survey: USA, France, Germany, Japan, China, Taiwan, Brazil, and Mexico). The government's role in USA was to develop high and effective infrastructure and leave it for private sectors [10]. However, Gibbs, Kraemer & Dedrick [14] research shows that "government regulation, such as privacy protection, can be critical to supporting e-commerce". Comparing developing countries to developed countries, the need for government incentive is high especially in Taiwan [15].

In regard to Saudi Arabia, the rapid growth of consuming ICT is notable. Saudi Arabia is the largest and fastest growth of ICT marketplaces in the Arab region [3], [4], [5], [6]. However, so far, the efforts towards e-commerce development in the country have not reached its originally stated aspirations; neither what it sees as the world's expectations of a country of the level of importance and weight in the global economy like Saudi Arabia [7], [8], [9]. In 2001, the Saudi Ministry of Commerce introduced a good step towards e-commerce development in the country by establishing a committee for e-commerce including members from different government and private sectors [3]. The roles of this committee were to follow the developments in the field of e-commerce and take the necessary steps to keep pace with them. They sought to identify the particular needs and requirements to take advantage of e-commerce, following-up to completion of the development work required, and the preparation of periodic reports on progress of work on a regular basis [3]. The committee has prepared a general framework for a plan to apply e-commerce systems in Saudi Arabia. This framework includes the improvement of various factors involved with e-commerce transactions (e.g. IT infrastructure, payment systems, security needs, legislations and regulations, delivery systems etc). The plan also includes the development of e-commerce education and training [3]. However, this information was gained from the first publication booklet of Saudi Ministry of Commerce in regards to e-commerce. No other information has been found either on Saudi Ministry of commerce website or its documents to provide further details about this committee and its current role in e-commerce development in the country. During Nov and Dec 2010, several e-mails were sent to the Saudi Ministry of Commerce to seek further information about e-commerce committee and gain its contact details. Unfortunately, no response was received. This followed by phone calls to the Ministry of Commerce to gain information for the same purpose. After several calls with different people in the Ministry, who have not heard about this committee, the answer was found with a person used to be involved with the activities of this committee. The researcher was told that the e-commerce committee no longer exists and the role of e-commerce supervision and development has been transferred to the Ministry of ICT since 2005. The researcher then followed his contacts with Ministry of ICT seeking further information about the role of the ICT Ministry in terms of supporting e-commerce in the country. It has been told, through e-mail contact, nothing significance has been done specifically for e-commerce and they are currently conducting a survey on e-commerce in Saudi Arabia and a report may be published in June 2011. Saudi Arabia ranked 52 in e-readiness in the latest report of the global e-readiness ranking which assessed the quality of 70 countries' ICT infrastructure and the ability of their government, businesses and people to use ICT [16].

Several studies have been conducted to figure out the issues behind the delay of e-commerce development in Saudi Arabia. While these studies are several years back and talk about e-commerce in general, they still good to look at them. The Internet connection, seed and cost were the main obstacle for individuals and organizations to be involved in such electronic activities [7], [8]. While Internet penetration increased to 41% of the population by the end of Q3 2010 in Saudi Arabia [17], Broadband subscriptions (12.2% of the population) still too low compared to the developed nations. Some inhibitors also involved resistance to social adaptation to a new commercial paradigm, lack of trust of online business, shortage of skilled employees for implementation and maintenance of e-business systems [5], [17].

From public and private Saudi organizations perspectives, the most significant obstacle to e-commerce in Saudi Arabia is the lack of individual house addresses and also the lack of clear regulations, legislation, rules and procedures on how to protect the rights of all involved parties [18]. Before 2005 individuals had no uniquely identifying home addresses and the mail was not delivered to homes and offices [19]. Nowadays, individual house addresses may not represent a problem because Saudi Post has approved the postal delivery to homes and buildings in 2005 [19], [6]. Today, every resident in the main cities in Saudi Arabia can contact Saudi Post and order registering his building for free service provided by Saudi Post. However, while this service is still relatively new, Saudi Arabia is very late in providing individual addresses. Problems with adopting this service might be the citizens' lack of awareness of this service or the importance of mailboxes, their ignorance of the direct addresses for their houses with numbers and streets names, or their mistrust of receiving their mail in this way. Consequently, more efforts are needed to motivate the citizens owning house mailboxes and solve the problems that they face.

Regarding the lack of clear e-commerce law in Saudi Arabia, this is not only mentioned by commercial organizations [18], [20], but also is raised as an issue by Saudi consumers [21], [22]. Although Saudi Arabia contributes to the efforts of UNCITRAL (United Nations Commission into International Trade Laws) [3], there is a need to have major development in terms of e-commerce regulations, legislations and rules to protect the rights of all parties involved in e-commerce transactions [7], [18].

In 2007, Saudi Communication and Information Technology Commission (CITC) carried out an extensive study to evaluate the current situation of the Internet, and various aspects involving the Internet usage in Saudi Arabia. One of these aspects is e-commerce awareness and activities. As introduced earlier, for the business commercial organizations, they have found that only 9% of Saudi commercial organizations, mostly medium and large companies from the manufacturing sector, are involved in e-commerce implementation. It is reported that only 4 out of 10 private companies have their own website. This percentage takes on a higher proportion for the larger oil, gas and manufacturing companies. For the users (customer), they have found that 43% of the respondents were aware of e-commerce and only 6% ever bought or sold products online, "mainly airline tickets and hotel bookings" [9].

A study was carried out based on a literature review to explore the status of e-commerce in the Gulf countries (Saudi Arabia, Oman, UAE, Kuwait, Qatar, and Bahrain) in relation to international e-commerce including. The possible issues affect the growth of e-commerce in these countries include ICT infrastructure; awareness and education levels, confidence in online transactions, law awareness for e-commerce, trust and security issues, usability and interactivity of websites, change management, IT skills development and language fluency, level of assurance and encouragement from government and online companies, promotions of banks of online payment systems, and industry standards and competitive advantage [23].

In a previous paper, AlGhamdi, Drew & Al-Gaith [24] adopted a qualitative approach and used information obtained from a series on interviews with 16 Saudi retailers to form a list of factors that inhibit or discourage retailers from adopting online retail. These inhibitors are (1) setup cost; (2) delivery issues; (3) resistance to change; (4) lack of e-commerce experience; (5) poor ICT infrastructure; (6) lack of online payment options to build trust; (7) do not trust online sales; (8) habit/culture of people in Saudi Arabia not favorable toward buying online; (9) lack of clear rules/law for e-commerce in Saudi Arabia; (10) difficulties in offering competitive advantage on the Internet; (11) not profitable/ not useful; and (12) type of business/products are not suitable to be sold online. This qualitative study also established a list of incentives that are likely to enable or encourage retailers to adopt the online channel. These enablers are (1) develop strong ICT infrastructure, (2) government support and assistance for e-commerce, (3) educational programs and building the awareness of e-commerce, (4) trustworthy and secure online payment options, and (5) provision of sample e-commerce software for trialling. The main purpose of the current paper is to obtain quantitative indications of the relative strengths of these inhibiting and enabling factors.

In analyzing factors on these two lists and on similar lists (e.g, those highlighted in the above literature review) the Diffusion of Innovations (DOI) model can serve as a useful conceptual framework. Rogers [2] defined an innovation as an idea, practice or object that is perceived as new by an individual or other unit of adoption, and diffusion as "the process during which an innovation is communicated through certain channels over time among members of a social system". DOI model has been widely used to explain the adoption of innovations, especially those involving technology [25], [26], [27].

The DOI model sees the innovation diffusion process as involving five aspects. These are (1) perceived attributes of the innovation which enable/inhibit its adoption, (2) type of innovation-decision (optional, collective, authority), (3) communication channel diffusing the innovation at various stages in the innovation-decision process (mass media, interpersonal), (4) nature of the social system (norms, degree of network interconnectedness, etc), and (5) extent of change agent's promotion efforts [2]. In turn, the five perceived attributes of an innovation are its relative advantage, compatibility with the status quo, complexity, trialability, and observability [2].

# 3. RESEARCH METHODOLOGY

The whole project studying online retail in Saudi Arabia is built on the combination of qualitative and quantitative approaches. Qualitative study conducted first for exploration purpose and followed by quantitative approach based on qualitative findings for testing purpose. This type of approach is called exploratory mixed methods design [28] which is done "to explore a phenomenon, and then [collect] quantitative data to explain relationships found in the qualitative results" [28]. The mixed methods approach helps to provide an in-depth investigation of the research problem [29], [30], [31], [32], [33].

While the qualitative study was conducted in a separate research [24], the focus of this paper is on quantitative analysis based on the qualitative findings. In the qualitative study [24], interviews were conducted with 16 retailers' decision makers (including owners, headquarter managers, marketing managers, and IT managers) covering various business categories in Saudi Arabia. A qualitative content analysis was used to identify the factors that positively and negatively influence retailers to adopt and use online channel for sale.

In this paper, a questionnaire survey based on the qualitative study's findings is used to gain more information about the relative strengths of these factors. Typically a question that asks for information about the participating retailer's attributes would provide a set of choices, plus an open answer (e.g., "other") where the participant could insert additional information if he or she wishes. The two key questions are "what factors inhibit or discourage your company from implementing an online system to sell on the Internet?" and "what factors help or encourage your company to implement an online system to sell on the Internet?" The participants are given a list of 13 options to select from for the former question, and 6 options for the latter (in each case, the last option is "other"). Respondents may select as many of the available options as they wish, including the open answer. The survey questions are designed in English with an Arabic translation version being available, so that the participant can opt for the most familiar language.

Around 200 paper copies of the questionnaire forms were distributed in person to retail businesses in Jeddah (the main economic city in Saudi Arabia), Riyadh (the capital city), and Al-Baha during February and March of 2011. Potential participants were selected via the "snowballing" approach where some participants were initially approached, then they would be asked to recommend others who might be willing to participate, and so on. Judgment was also exercised to ensure that the questionnaire forms were delivered to a wide range of businesses in terms of their size, and the type of products or services that they offered. A total of 80 completed forms were returned, giving a response rate of around 40%.

Electronic copies of the questionnaire forms were also kept on the website of the Griffith University research survey center (English version / Arabic version). The authors collected the email addresses of 416 retail companies which were members of the Jeddah and Riyadh chambers of commerce. Associates of the authors provided an additional list of around 50 business email addresses with which they were familiar. Invitations to participate online were sent via email to these 466 addresses, but around 100 were returned because the addresses were invalid. The authors also received Saudi Post's assistance in emailing the 50 retailers which had registered with their e-mall (this e-mall had started in October 2010). Thus, in total there were 416 retailers who potentially could choose to participate in the online survey via the Griffith website. At the time of writing (May 2011), 68 had done so, implying a response rate of 16.3%.

## 4. RESULTS AND DATA ANALYSIS

This section presents a summary and analysis of the responses collected to date from 148 participants. Small retailers represent 32.4% of the sample, compared with 41.2% for medium-sized and 26.4% for large companies. About 75.7% of the businesses in the sample use computers as primary, essential tools, whist 16.9% use them only as secondary tools, and 7.4% do not use them at all. Only 55.4% of the participating retailers have a website.

Given that nearly one-half (44.6%) of the participants do not own a website, it is not surprising that a large majority (71.6%) of all retailers in the sample report having never conducted online sales. As for the businesses (28.4%) which have done so, online sales represent less than 10% of total sales for approximately one-third of this group (9.5% of total sample). For another one-third of the group (8.8% of the sample) the online-to-total sales ratio is between 10 and 50%, with the remaining one-third (10.1% of sample) reporting online-to-total sales ratios over 50%. These statistics are consistent with previous findings indicating that the online retailing industry in Saudi Arabia has not developed very rapidly or vigorously.

Table 1 presents, for each category of participating businesses, the percentage of those which have adopted the online sale channel. Rather surprisingly, a higher percentage of small businesses (28.4%) than medium-sized (21.3%) or large businesses (25.6%) have done so. The rates of adoption of online retail are noticeably lower among businesses selling building materials (0.0%), autos, parts and accessories (0.0%), and consumer electronics (9.1%). By contrast, adoption rates are clearly high among vendors of perfume and beauty products (69.2%), and businesses which already have a website (39.0%). The results for some of the other categories may appear highly interesting but often there are too few respondents in each category (e.g., sporting goods) and so these results must be treated with some caution.

**Table 1**. Company attributes and rate of adoption of online sale channel

| **Category of retailers** | **No. of retailers** | **Have sold online** | |
|---|---|---|---|
| | | **Number** | **%** |
| All participating retailers | 148 | 42 | 28.4 |
| *Company size* | | | |
| Small | 48 | 19 | 39.6 |
| Medium | 61 | 13 | 21.3 |
| Large | 39 | 10 | 25.6 |
| *Product or service type* | | | |
| Apparel, accessories, and footwear | 16 | 4 | 25.0 |
| Appliances and home improvement | 18 | 5 | 27.8 |
| Books and school needs | 4 | 1 | 25.0 |
| Building materials | 8 | 0 | 0.0 |
| Cars, auto parts, and accessories | 8 | 0 | 0.0 |
| Computer-related | 8 | 2 | 25.0 |
| Consumers electronics | 11 | 1 | 9.1 |
| Furniture | 7 | 2 | 28.6 |
| Groceries | 17 | 4 | 23.5 |
| Jewellery | 1 | 0 | 0.0 |
| Medicines and medical equipments | 4 | 0 | 0.0 |
| Optical products | 2 | 0 | 0.0 |
| Perfumes and beauty products | 26 | 18 | 69.2 |
| Printing equipment and/or services | 2 | 0 | 0.0 |
| Sporting goods | 2 | 2 | 100.0 |
| Telecommunications services | 2 | 1 | 50.0 |
| Toys and video games | 3 | 1 | 33.3 |
| Travel and tourism | 4 | 0 | 0.0 |
| Other | 5 | 1 | 20.0 |
| *Website* | | | |
| Business already has a website | 82 | 32 | 39.0 |
| | | | |

Table 2 reports our findings with respect to the relative importance of factors that inhibit retailers in Saudi Arabia from adopting e-commerce. In this table, the inhibitors are listed in the order in which they are presented to the respondents. Figure 1 illustrates the same information, but with each inhibitor being ranked according to its relative weight, from being most frequently selected to least.

**Table 2**. Inhibitors of adoption by Saudi retailers of the online channel

| Identifier | Inhibitor | Selected by % of respondents | Rank |
|---|---|---|---|
| IN1 | Setup cost | 10.1 | 10 |
| IN2 | Cannot offer delivery service | 10.8 | 9 |
| IN3 | Not trusting online sales activities | 12.8 | 8 |
| IN4 | Not be profitable/ useful for us | 8.1 | 11 |
| IN5 | Products are not suitable to be sold online | 25.0 | 3 |
| IN6 | Cannot offer a competitive advantage over competitors | 7.4 | 12 |
| IN7 | Resistance to change | 16.9 | 6 |
| IN8 | Lack of clear legislations and rules of e-commerce in KSA | 35.8 | 2 |
| IN9 | Lack of e-commerce experience | 35.8 | 2 |
| IN10 | Current habits of people in KSA does not suit online transactions | 42.6 | 1 |
| IN11 | Lack of online payment options in KSA to help build the trust of customers | 20.9 | 5 |
| IN12 | Poor ICT infrastructure | 22.4 | 4 |
| IN13 | Others | 13.6 | 7 |

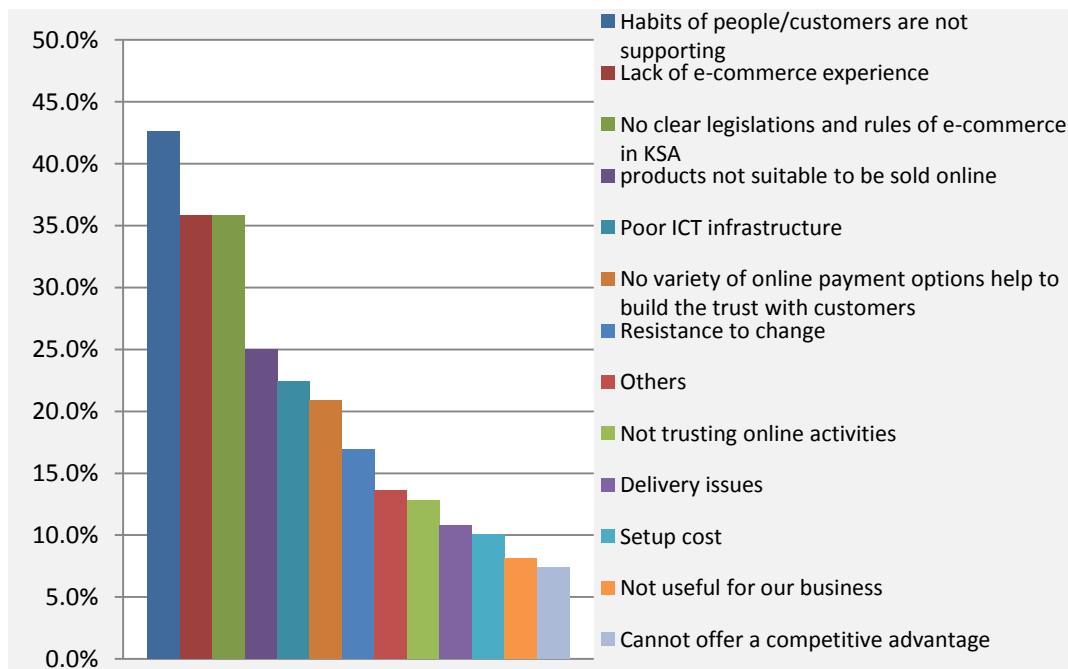

**Figure1.** Factors inhibiting adoption of online retailing by Saudi vendors

From Table 2 and Figure 1, it can be seen clearly that some of the most serious inhibitors are outside the direct control of retailers. For example, ordinary action by individual retailers is unlikely to have much weight in countering IN10 (Current habits of people in KSA do not suit online transactions, ranked 1 by retailers), IN8 (Lack of clear legislations and rules of e-commerce in KSA, ranked 2), IN5 (Products are not suitable to be sold online, ranked 4), IN12 (Poor ICT infrastructure, ranked 5) and IN11 (Lack of online payment options in KSA to help build the trust of customers, ranked 6). Nevertheless, there are some important inhibitors over which retailers do have some control, such as IN9 (Lack of e-commerce experience, ranked 2), IN7 (Resistance to change, ranked 7).

Table 3 and Figure 2 present basic survey results with regard to factors that tend to enable or encourage Saudi businesses to engage in online retailing.

**Table 3**. Enablers of adoption by Saudi retailers of the online channel

| Identifier | Enabler | Selected by % of respondents | Rank |
|---|---|---|---|
| EN1 | Develop strong ICT Infrastructure | 39.9 | 3 |
| EN2 | Provision of sample e-commerce software for trialling | 25.7 | 5 |
| EN3 | Government support and assistance for e-commerce | 53.4 | 2 |
| EN4 | Providing trustworthy and secure online payment options | 58.1 | 1 |
| EN5 | Educational programs for people and building the awareness of e-commerce in the country | 31.1 | 4 |
| EN6 | Others | 15.5 | 6 |

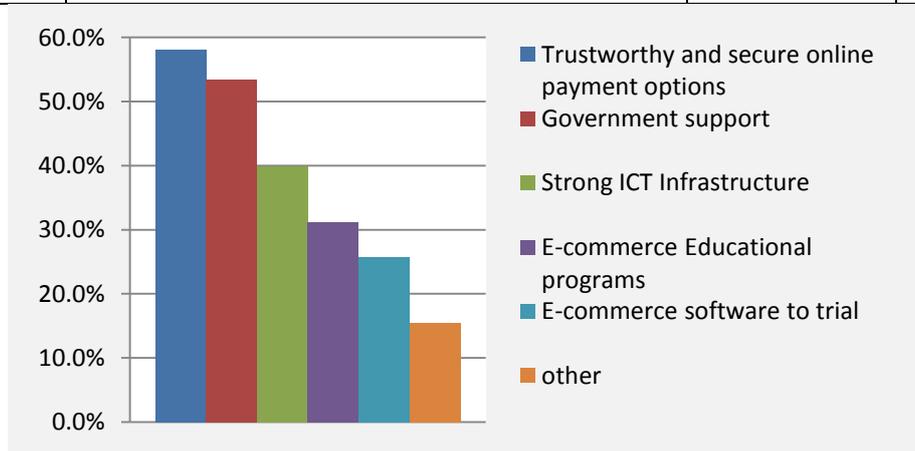

**Figure2.** Factors facilitating adoption of online retailing by Saudi vendors

The top enablers from the viewpoint of retailers are all dependent on government action, either directly or indirectly. They include EN4 (Provision of trustworthy and secure online payment options, ranked 1), EN3 (Government support and assistance for e-commerce, ranked 2), EN1 (Development of strong ICT Infrastructure, ranked 3) and EN5 (Educational programs for people and building the awareness of e-commerce in the country, ranked 4).

Delving more deeply into the details of the individual responses allows us to gain further insights into key inhibitors and enablers of the decision by Saudi retailers to adopt the online channel. Table 4 presents data relating to the interactions between vendor attributes and inhibitors, and Table 5 does the same for enablers. For example, Table 4 confirms that IN1 (Setup cost) is of little concern to retailers which currently use computers as primary (essential) tools, even though it is a major inhibitor from the viewpoint of retailers who use them only as secondary tools or not at all. Similarly, IN8 (Lack of clear regulations) is selected by only 21.4% of the retailers which have conducted online sales; the corresponding figure for retailers which have not done so is 38.8% (almost double). It is significant that, for both groups, the most serious inhibitor is IN10 (Habits of the Saudi public), being selected by more than 40% of the participating businesses in each group.

**Table 4**: Inhibitors and some indicators of e-commerce readiness among the sample

|  |  |  | Percentage of respondents selecting |  |  |  |  |  |  |  |  |  |  |  |
|---|---|---|---|---|---|---|---|---|---|---|---|---|---|---|
|  |  | % of sample | IN1 | IN2 | IN3 | IN4 | IN5 | IN6 | IN7 | IN8 | IN9 | IN10 | IN11 | IN12 |
| Computer Use | Primary | 75.7 | 8.9 | 9.8 | 11.6 | 7.1 | 22.3 | 6.3 | 17.9 | 37.5 | 33.0 | 45.5 | 17.9 | 20.5 |
|  | Secondary | 16.9 | 88.0 | 12.0 | 20.0 | 12.0 | 32.0 | 8.0 | 12.0 | 36.0 | 44.0 | 36.0 | 36.0 | 28.0 |
|  | Not using | 7.4 | 81.8 | 18.2 | 9.1 | 9.1 | 36.4 | 18.2 | 18.2 | 18.2 | 45.5 | 27.3 | 18.2 | 27.3 |
| Website | Yes | 55.4 | 4.9 | 11.0 | 9.8 | 8.5 | 20.7 | 2.4 | 13.4 | 37.8 | 29.3 | 43.9 | 22.0 | 19.5 |
|  | No | 44.6 | 16.7 | 10.6 | 16.7 | 7.6 | 30.3 | 13.6 | 21.2 | 33.3 | 43.9 | 40.9 | 19.7 | 25.8 |
| Selling Online | Yes | 28.4 | 14.3 | 9.5 | 0 | 7.1 | 4.8 | 7.1 | 2.4 | 21.4 | 14.3 | 40.5 | 19.1 | 14.3 |
|  | No | 66.2 | 9.2 | 12.2 | 19.4 | 9.2 | 35.7 | 8.2 | 24.5 | 38.8 | 45.9 | 44.9 | 23.5 | 25.5 |

**Table 5**: Enablers and some indicators of e-commerce readiness among the sample

|  |  |  | Percentage of respondents selecting |  |  |  |  |
|---|---|---|---|---|---|---|---|
|  |  | % of sample | EN1 | EN2 | EN3 | EN4 | EN5 |
| Computer Use | Primary | 75.68 | 46.43 | 25.89 | 55.36 | 57.14 | 33.93 |
|  | Secondary | 16.89 | 16.00 | 16.00 | 60.00 | 68.00 | 24.00 |
|  | Not using | 7.43 | 27.27 | 45.45 | 18.18 | 45.45 | 18.18 |
| Website | Yes | 55.41 | 50.00 | 25.61 | 62.20 | 64.63 | 29.27 |
|  | No | 44.59 | 27.27 | 25.76 | 42.42 | 50.00 | 33.33 |
| Selling Online | Yes | 28.38 | 23.81 | 11.90 | 21.43 | 47.62 | 14.29 |
|  | No | 66.22 | 43.88 | 31.63 | 63.27 | 60.20 | 37.76 |

The last two rows of Table 5 present an interesting contrast. About two-thirds (66.2%) of the respondents have not had online sales experiences, and they tend to rate the nominated enablers quite highly: each of the 5 enablers is selected by 31% or more of these participants, and 2 enablers are selected by over 60% of them. To some extent, these high ratings reflect some unrealistic expectations about the true efficacy of the enablers. Responses from the 28.4% of the sample which do have actual experience with operating an online retail business suggest that the listed enablers are no magic bullets. EN3 (government support), for example, is selected by only 21.4% of these retailers. From their viewpoint, the top-rated enabler is EN4 (Provision of trustworthy and secure online payment options).

## 5. THEORETICAL AND PRACTICAL IMPLICATIONS

The findings highlighted in the previous section are consistent with the DOI theoretical framework. In particular, the lists of top inhibitors and enablers of the decision by Saudi retailers to adopt the online channel feature prominently items which relate to the attributes of this innovation, such as its relative advantage compared with traditional retail channels, its compatibility with the potential adopters' existing situation, the observability of successes achieved by early adopters, etc. Further, the nature of the social system (which in this case tends to raise expectations that the government will assume the role of key agent of change) and the extent of promotion efforts (or lack of them) on the part of change agents are clearly a part of the underlying story to date.

The most relevant practical implications of this paper are probably those which can be drawn from the responses of Saudi businesses which already have experience making online sales to Saudi customers. According to them, the most serious inhibitors are IN10 (Unfavourable Saudi consumer habits), IN8 (Lack of government regulations), and IN11 (Lack of online payment options). Of these, IN10 is a long-standing condition and will only change gradually, if at all. By contrast, the government and the industry are in a position to affect IN8 and IN11 directly.

Retailers with previous online sale experience also provide clear and practical indications as to the top enabling factors. These are EN4 (Developing online payment options), EN1 (Enhancing ICT infrastructure) and EN3 (Government regulations and support). Of these EN4 was selected by 48% of these retailers, compared with 24% for EN1 and 21% for EN3. As the issue of online payment options is given such emphasis, it may be useful to examine it in some detail.

In the West, using credit cards to pay is the most popular method to conduct online purchases. In Saudi Arabia, however, many consumers are reluctant to use credit cards, both because of a lack of trust and because some consumers are culturally averse to carrying out transactions linked with conventional interest rates. Thus, providing alternative, trustworthy and easy-to-use payment options is a critical need for the industry. Possible solutions include debit cards and payment systems such as PayPal. Another option is the electronic bill presentment and payment system which Saudi Arabia has developed for billers and payers who are resident in the country, called SADAD. In essence, the system facilitates data exchange between registered billers and the nation's commercial banks, and relies on existing banking channels (such as internet banking, telephone banking, ATM transactions, and even counter transactions) to allow bill payers to view and pay their bills via their banks (for more details, see [34]). Many consumers are comfortable with using SADAD. However, follow-up comments from some of the respondents in our study suggest that small-to-medium businesses see the initial costs of registration with SADAD and the ongoing transactions processing fees as being too high.

# 6. CONCLUSION

This paper presented some findings from a study researching the diffusion and adoption of online retailing in Saudi Arabia. It identified and explored key factors influence Saudi retailers to adopt e-commerce selling to their customers. While the overall research project uses mixed methods, the focus of this paper was on a quantitative analysis of responses obtained from a survey of retailers in Saudi Arabia, with the design of the questionnaire instrument being based on the findings of a qualitative analysis reported in a previous paper. From the study findings, it can be seen clearly that the most serious inhibitors are the current habits of people in KSA do not suit online transactions, lack of clear legislations and rules of e-commerce in KSA and lack of e-commerce experience. On the other hand, the top enablers from the viewpoint of retailers are all dependent on government action, either directly or indirectly. They include provision of trustworthy and secure online payment options (ranked 1), government support and assistance for e-commerce (ranked 2), development of strong ICT Infrastructure (ranked 3) and educational programs for people and building the awareness of e-commerce in the country (ranked 4). However, responses from the 28.4% of the sample which do have actual experience with operating an online retail business suggest that high ratings reflect some unrealistic expectations about the true efficacy of the enablers.

The most relevant practical implications of this paper are probably those which can be drawn from the responses of Saudi businesses which already have experience making online sales to Saudi customers. According to them, the most serious inhibitors are unfavourable Saudi consumer habits, lack of government regulations, and lack of online payment options. Retailers with previous online sale experience also provide clear and practical indications as to the top enabling factors. These are developing online payment options, enhancing ICT infrastructure and government regulations and support. Therefore, the government and the industry should pay attentions to these factors to facilitate e-retail growth in the country. The critical question, then, becomes whether there are valid justifications for the government to take such an interventionist role in normal commerce, as opposed to the cases of e-government and e-learning which involve public services or "social" goods. Such a question must be left to future research.